\documentclass[
 aip,
 amsmath,amssymb,
 reprint,
]{revtex4-1}

\usepackage{graphicx}
\usepackage{dcolumn}
\usepackage{bm}
\usepackage[utf8]{inputenc}
\usepackage[T1]{fontenc}
\usepackage{mathptmx}
\usepackage{etoolbox}
\usepackage{gensymb}
\makeatletter
\def\@email#1#2{%
 \endgroup
 \patchcmd{\titleblock@produce}
  {\frontmatter@RRAPformat}
  {\frontmatter@RRAPformat{\produce@RRAP{*#1\href{mailto:#2}{#2}}}\frontmatter@RRAPformat}
  {}{}
}%
\makeatother

\makeatletter
\DeclareRobustCommand{\textsupsub}[2]{{
  \m@th\ensuremath{
    ^{\mbox{\fontsize\sf@size\z@#1}}
    _{\mbox{\fontsize\sf@size\z@#2}}
  }
}}
\makeatother
\begin{document}

\preprint{AIP/123-QED}

\title[Three-Electrode Cell Calorimeter for Electrical Double Layer Capacitors]{\textbf{Three-Electrode Cell Calorimeter for Electrical Double Layer Capacitors}}
\author{Joren E. Vos}
\author{Hendrik P. Rodenburg}
\author{Danny Inder Maur}
\author{Ties J. W. Bakker}
\affiliation{ 
Van ’t Hoff Laboratory for Physical and Colloid Chemistry, Debye Institute for Nanomaterials Science, Utrecht University, Padualaan 8, 3584 CH Utrecht, The Netherlands
}
\author{Henkjan Siekman}
\affiliation{\mbox{Instrumentation Department, Utrecht University, Sorbonnelaan 4, 3584 CA Utrecht, The Netherlands}
}
\author{Ben H. Erné}
\affiliation{ 
Van ’t Hoff Laboratory for Physical and Colloid Chemistry, Debye Institute for Nanomaterials Science, Utrecht University, Padualaan 8, 3584 CH Utrecht, The Netherlands
}

\date{\today}

\begin{abstract}
A calorimeter was built to measure the heat from a porous capacitive working electrode connected in a three-electrode configuration. This makes it possible to detect differences between cathodic and anodic heat production. The electrochemical cell contains a large electrolyte solution reservoir, ensuring a constant concentration of the salt solution probed by the reference electrode via a Luggin tube. A heat flux sensor is used to detect the heat, and its calibration as a gauge of the total amount of heat produced by the electrode is done on the basis of the net electrical work performed on the working electrode during a full charging-discharging cycle. In principle, from the measured heat and the electrical work, the change in internal energy of the working electrode can be determined as a function of applied potential. Such measurements inform about the potential energy and average electric potential of ions inside the pores, giving insight into the electrical double layer inside electrode micropores. Example measurements of the heat are shown for porous carbon electrodes in aqueous salt solution.
\end{abstract}

\maketitle

\section{\label{sec:intro}Introduction}
Capacitive porous electrodes are of interest for instance as supercapacitors in power delivery systems\cite{Simon2020} and as reversible salt absorbants in water desalination.\cite{Wang2021} During charging of a porous electrode, electrical energy and ions are stored in the electrical double layer (EDL). Experimental characterization of the EDL helps to elucidate the energetic or ionic uptake capacity of the electrode. Changes in the amount of charge can be measured in the external electrical circuit.\cite{Porada2012,Kim2015} Additional information on the charging mechanism and the amounts of ions inside the pores can for instance be obtained from in situ NMR spectroscopy,\cite{Forse2016,Griffin2015,Forse2015,Wang2013} infrared spectroscopy,\cite{Richey2013,Richey2014} and small-angle neutron scattering.\cite{Boukhalfa2013,Boukhalfa2014} Here, we will focus on a thermodynamic characterization approach that consists of measuring the heat exchanged while the electrode is being charged or discharged.

Electrodes in any electrochemical cell produce heat, although this is generally not the intended outcome. One example is heat generation during electrolysis reactions.\cite{Narger1990} Another is Joule heat produced by supercapacitors,\cite{Dandeville2011} which can cause a strong temperature rise that can be damaging for their performance.\cite{Hansen1982,Chen2016,AlSakka2009} When it is possible to determine the reversible heat, this provides valuable information on the change in thermodynamic state of the system. The reversible heat can for example correspond to the change in enthalpy of electrochemical reactions \cite{Turner1978,Sherfey_1958} or to the entropic heat from batteries, in agreement with the temperature dependence of their open circuit voltage.\cite{CHENG2022} For supercapacitors, the reversible heat has been interpreted in different ways, as the entropic heat from the confinement of ions into the pores of the electrodes,\cite{Schiffer2006} or as changes in the entropic part of the grand potential energy,\cite{Janssen20172,Glatzel2021} or as due to several entropic and enthalpic contributions because of mixing as well as electrical and steric interactions of the ions,\cite{DEntremont2014,DEntremont2015,Kundu2021} or as due to nonzero potential energy of the ions in the pores.\cite{heatmanuscript}

Measuring heat from porous electrodes requires a different measurement approach than measuring heat from submonolayer changes at a flat electrode,\cite{Schuster2017} which result in very little heat, produced very briefly.\cite{Etzel2010} This requires highly sensitive and rapid detection, which can for instance be achieved \mbox{using} lithium tantalate-based sensors.\cite{Frittmann2015} Porous electrodes have a much higher surface area and slow ionic transport in an extensive porous network,\cite{Ma2022,Lian2020} resulting in much more heat production but spread out over a much longer time. Due to the long duration of heat production, the measurement requires a very stable background temperature to differentiate from heat exchange due to temperature changes in the environment.

Here, a setup is presented that measures the heat of charging and discharging from a capacitive porous carbon\cite{Sakintuna2005} electrode, connected in a 3-electrode configuration. The setup was first used in Ref.~\onlinecite{heatmanuscript}, where it was described much more briefly. Earlier experiments on capacitive porous electrodes were done on 2-electrode cells, by measuring the temperature of the complete cell using a resistance temperature detector,\cite{Janssen2017,Schiffer2006} or by measuring the separate heats of both electrodes, using heat flux sensors.\cite{Munteshari2018,DEntremont2015} When the heat of a complete cell is measured, differences between cathodic and anodic heat production cannot be distinguished. This limitation disappears when the heat of individual electrodes is measured. However, when the cell has only two electrodes, even though it is clear that the charge that exits one electrode enters the other electrode, it is more difficult to clarify differences between cathodic and anodic behavior, because the potentials applied to each electrode are not determined against a constant reference. In the setup presented here, a reference electrode is introduced as the third electrode. The current still flows from the working electrode to a counter electrode, but the potential on the working electrode is applied and measured with respect to an \mbox{invariant} reference electrode. A three-electrode cell is commonly used in electrochemistry,\cite{Faulkner2001} but not for measurements on commercial batteries or supercapacitors. In Section~\ref{sec:experimental}, the design and operation of the setup are presented, and typical measurements are shown in Section~\ref{sec:results}. 

\section{\label{sec:experimental}Design and Operation}
\subsection{Electrochemical setup}
\begin{figure}
\includegraphics[scale=0.75]{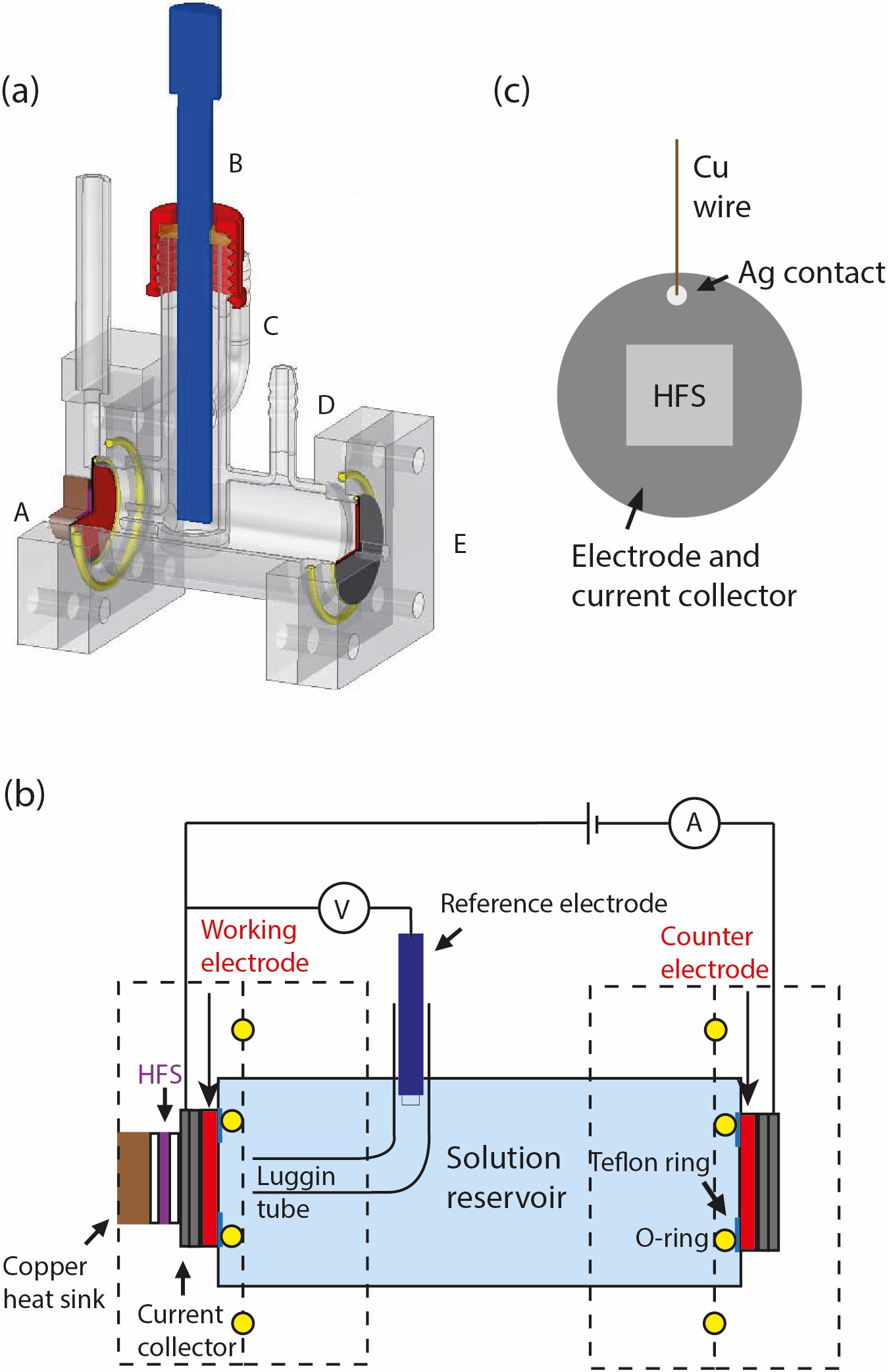}
\caption{\label{fig:cell}(a) Technical drawing of the electrochemical cell used for heat measurements on a porous electrode, connected in a 3-electrode configuration.
A 3D pdf of this figure is provided in the Supporting Information; A = outer square part containing the working electrode, with the HFS in purple  behind it; the HFS is thermally stabilized by a copper block at the back; B = reference electrode; C \& D = electrolyte solution in- and outlets; E = outer square part containing the counter electrode. (b) Schematic overview of the cell and its connections to the potentiostat. The yellow circles are cross-sections of the O-rings seen in part (a) of the figure. (c) Position of the HFS with respect to the electrode, the current collector, and the ohmic contact.}
\end{figure}

The electrochemical cell developed to measure heat effects of capacitive porous electrodes in a 3-electrode configuration is shown in Fig.~\ref{fig:cell}. The cell has three glass parts. The central part consists of a horizontal cylinder (6.4\,cm in length, 2.5\,cm in external diameter) whose extremities are glued into the central hole (2.5\,cm diameter) of square blocks (5\,cm by 5\,cm, 5\,mm thick). These glued square blocks of the central part are connected to two outer square parts of the cell via plastic screws inserted into four holes at the corners of the squares, see Fig.~\ref{fig:cell}(a). One of the outer square parts contains the counter electrode (CE), and the other contains the working electrode (WE) and the heat flux sensor (HFS). The WE and CE are mounted vertically, allowing gas to escape from the electrode surface. The separation of 6.4\,cm between the electrodes ensures that no measurable heat of the CE reaches the heat flux sensor mounted behind the WE. The volume of the cylinder (30\,mL) is sufficiently large that the salt concentration remains approximately constant. The reference electrode (RE) senses the potential of the solution near the WE via a Luggin tube. Typically, a Radiometer Analytical REF201 Red Rod Ag/AgCl/saturated KCl is used as RE.

The WE and the CE each consist of a disk of porous carbon with a diameter of 22\,mm and a thickness of typically 0.4\,mm. Compared to a Pt CE, a porous carbon CE has the advantage that it does not produce hydrogen or oxygen gas under our measurement conditions, gases which can be oxidized or reduced at the WE, leading to faradaic currents which complicate the interpretation. At the center, these electrodes are glued to a nonporous carbon disk of 25\,mm in diameter using a minimal amount of nonconductive Bison Kombi Snel epoxy glue. Mechanical contact between the WE and the current collector (a nonporous carbon disk of the same dimensions) is realized by pushing them together at their outer rims with a flat 50\,µm thick Teflon ring itself pushed by a 2\,mm thick rubber O-ring with a diameter of 21\,mm. The central part of the electrode exposed to the solution has a diameter of 18.5\,mm. An electrically insulated copper wire is glued to the back side of the current collector using silver epoxy glue (Chemtronics\textsuperscript{®} CW2400 conductive epoxy), ensuring electrical contact. This is topped off with nonconductive epoxy glue to insulate electrically the outer portion of the silver epoxy glue. The HFS (greenTEG gSKIN\textsuperscript{®} XP 26 9C, earlier used to measure heat from supercapacitor electrodes by Munteshari et al.\cite{Munteshari2018}), is placed behind the WE in a separate glass compartment with walls of 0.15 mm in thickness. This compartment is an additional protection of the sensor (which must remain dry) against salt solution leaking around the outer rim of the current collector. The 1\,cm\,$\times$\,1\,cm surface of the HFS faces the electrode and current collector and is centered with respect to them. In contrast, the ohmic contact between current collector and copper wire is more to the side, see Fig.~\ref{fig:cell}(c).

The HFS voltage is sampled twice per second using a Keithley 2182A Nanovoltmeter, connected to a personal computer via a GPIB interface. The electric potential is applied between the WE and the CE using a channel of an AMETEK PARSTAT MC-1000 multichannel potentiostat, with a feedback loop to keep the electrical potential of the WE stable with respect to the RE. The same instrument measures the resulting current between the WE and the CE.

\subsection{Temperature stabilization}
The electrochemical cell is in a controlled thermostatic environment to ensure a stable background level of the heat flux signal. The HFS is in thermal contact with a copper heat sink of 5.5\,mm in thickness and 15.3\,mm in diameter, in contact with water that surrounds the electrochemical cell. The cell is submerged in a 2\,L glass beaker, filled with tap water up to the breathing hole of the reference electrode. This beaker is positioned at the center of a copper cylinder (34\,cm height by 24 cm width), itself at the center of a thermostated box of 50\,cm by 50\,cm by 45\,cm in height. Copper tubing is welded to the copper cylinder for good heat contact, and connected to a Julabo F25-HE Refrigerated/heating circulator bath via PVC tubes, pumping thermostated water through the copper cylinder. The box is closed with a lid measuring 50\,cm in length and width and 24\,cm in thickness and consisting of wood and styrofoam. The large heat capacity of the water in the glass beaker and an air gap of 5\,cm between the beaker and the copper walls dampen temperature fluctuations of the thermostatic bath. To verify temperature stability and independence from environmental artifacts, the temperatures of the water in the glass beaker, of the water in the thermostatic bath, and of the air in the room are monitored using Pt100 sensors, whose signals are acquired via a Pico\textsuperscript{®} Technology PT-104 Platinum Resistance Data Logger. The water bath has a temperature range of 5\,\degree{C} to 70\,\degree{C}. Due to the large heat capacity, it takes up to 48 hours before the thermostatic box reaches a stable temperature after closing the lid. Typically, the temperature of water in the glass beaker is constant within a standard deviation of 0.0045\,\degree{C} on a short time scale ($10^2$\,s) and 0.0058\,\degree{C} on a long time scale ($10^5$\,s).

\subsection{Operation and Calibration}
The standard measurement protocol is the same as in Ref.~\onlinecite{heatmanuscript}. The temperature is stabilized and the difference between the potential of the WE and that of the RE at open circuit—the open circuit potential ($V\textsubscript{OCP}$)—is measured for several hours, until this potential has also stabilized, at around 0.2\,V vs. RE in the case of the studied carbon electrodes in aqueous NaCl solution. A cyclic voltammogram is measured to verify that the electrical behavior of the WE is largely capacitive, without oxidative or reductive peaks indicating faradaic processes. Then, the same potential as was measured at equilibrium at open circuit, $V$\textsubscript{OCP}\,=\,0.2\,V vs. RE, is applied actively to the WE for $\geq$1\,hour until the current becomes minimal ($<$20\,µA). The applied potential is then changed by $\Delta{V}$, resulting in peaks of electrical current and HFS voltage, $V$\textsubscript{HFS}. After 1 to 2 hours, \,0.2\,V vs. RE is reapplied. The charging-discharging cycle is repeated for $\Delta{V}$\,=\,$+$0.5\,V to $+$0.1\,V and $\Delta{V}$\,=\,$-$0.5\,V to $-$0.1\,V vs. RE, and this potential series is repeated at least once.

In principle, calibration can be done based on the specifications of the HFS (12.7 \textmu{V} per heat flux in W/m\textsuperscript{2} for the sensor used here). In that case, it must be known which fraction of the heat produced by the electrode is measured by the HFS. This can be calculated by assuming that half of the electrode heat is produced in the direction of the HFS and by taking into account the different surface areas of the electrode and the HFS. However, differences in the distance between electrode and HFS in early prototypes of the cell, as well as differences in heat contact from one electrode to the other, can render the calibration unreliable by a factor of order 2. Therefore, we chose to calibrate on another basis: the Joule heat of a complete charging-discharging cycle.\cite{heatmanuscript}

When a known potential $\Delta{V}$ is applied vs. $V\textsubscript{OCP}$, a charge $\Delta{Q}$ builds up in the electrode during charging. As explained in Ref.~\onlinecite{heatmanuscript}, for a full cycle of charging and discharging, the total reversible heat is zero and all the measured heat is Joule heat, equal to the net electrical work performed on the electrode during the cycle:
\begin{equation}
q\textsubscript{ch}+q\textsubscript{dis}=-\Delta{Q}\Delta{V},
\label{eq:five}
\end{equation}
where $q\textsubscript{ch}$ is the heat during charging and $q\textsubscript{dis}$ the heat during discharging. This equation stems from the knowledge that the internal energy of the electrode is the same before and after a full cycle of charging and discharging, since the electrode comes back to its initial state. The total integrated surface area of the two HFS voltage peaks in a charging-discharging cycle is proportional to $-\Delta{Q}\Delta{V}$ and the ratio of the two gives the calibration constant $K$, the total amount of heat produced by the electrode per integrated HFS signal:
\begin{equation}
-\Delta{Q}\Delta{V}=K\int{V\textsubscript{HFS}\,dt}
\label{eq:two}
\end{equation}
Separate heats of charging and discharging in energy units can now be calculated from integrated HFS signals in units of Vs, using the calibration constant in units of J/(Vs). This assumes that the HFS has the same sensitivity for the reversible heat and for the Joule heat, an assumption which will be discussed in the next section.

\section{\label{sec:results}Test Measurements}
\subsection{Experimental Results}
\begin{figure}
\includegraphics[scale=0.33]{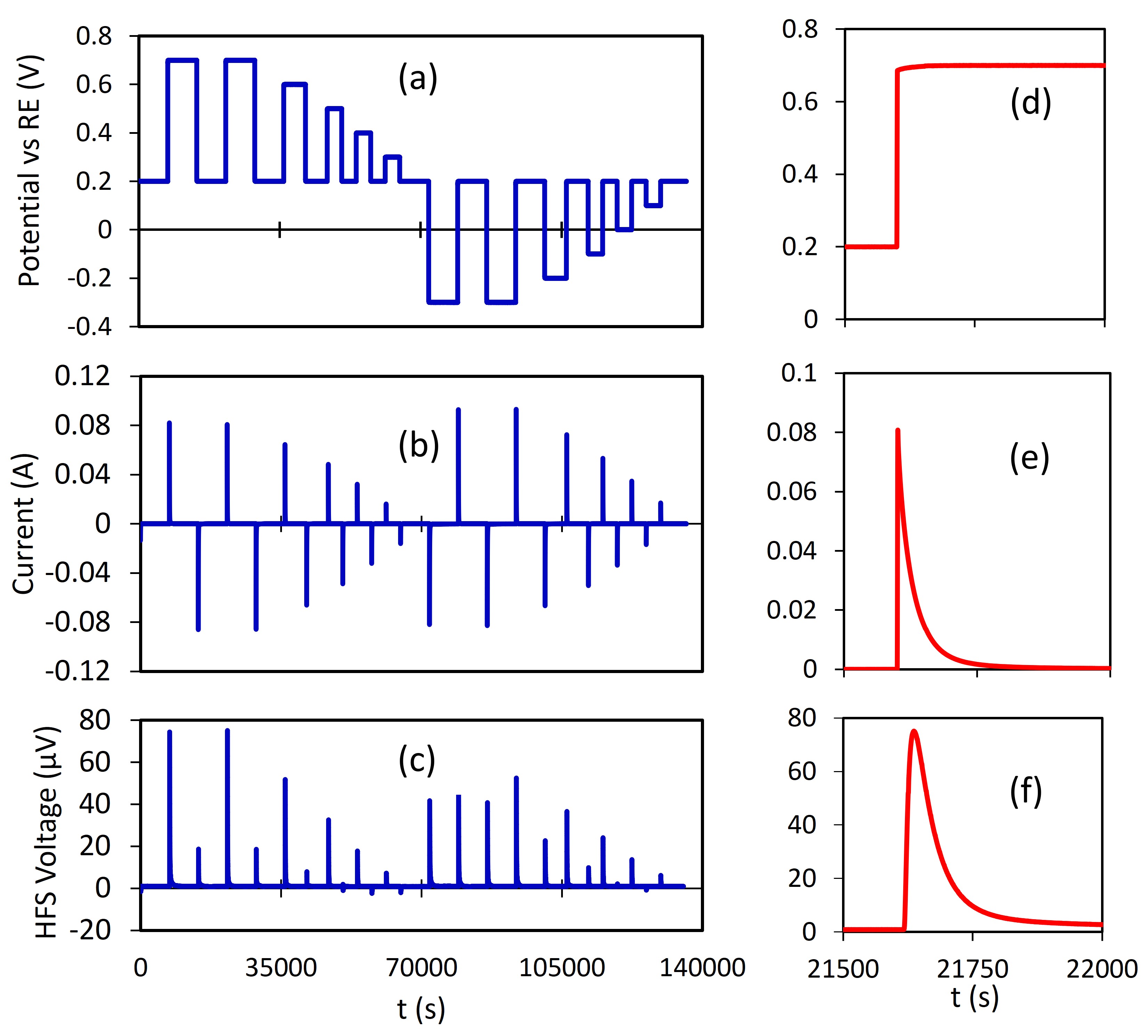}
\caption{\label{fig:HFS}Typical measurements on a porous carbon electrode in 1\,M aqueous NaCl with (a) the measured applied potential, (b) the resulting current, and (c) the HFS voltage. (d)-(f) Shape of the signals in (a)-(c), respectively, during charging from 0.2\,V to 0.7\,V (potentials versus RE; 0.2\,V is the open circuit potential).}
\end{figure}
Test measurements were done on porous carbon electrodes cut from sheets of electrode material produced by Voltea B.V. (Amstelveen, The Netherlands). The material had a density of 0.58\,gmL\textsuperscript{$-$1}, a porosity of 65\%, and a Brunauer-Emmett-Teller surface area of 88\,m\textsuperscript{2} per electrode.\cite{heatmanuscript} This material was comparable to that prepared in Ref.\onlinecite{Kim2015}, with activated carbon (YP-50F, Kuraray, Japan), carbon black (Vulcan XC72R, Cabot Corp., Boston, MA) and a binder (85:5:10 in weight ratio). The electrolyte solution consisted of 1\,M NaCl (for molecular biology, $\geq$98\%, Sigma), degassed using a Branson 8800 Series Ultrasonic Cleaner.

In Fig.~\ref{fig:HFS}, results of a typical measurement series are shown. Within minutes of changing the potential applied to the working electrode, current and heat flux were close to zero, but to ensure that equilibrium was attained and to have a reliable baseline, measurements were continued for at least one hour after each change of applied potential. The signal-to-noise ratio of the highest HFS voltage peak (75\,\textmu{V}) was 2850:1 in Fig.~\ref{fig:HFS}. In general, the signal-to-noise ratio was about two orders of magnitude higher than in the 2-electrode thermometer setup of Ref.~\onlinecite{Janssen2017}.

\begin{figure}
\includegraphics[scale=0.35]{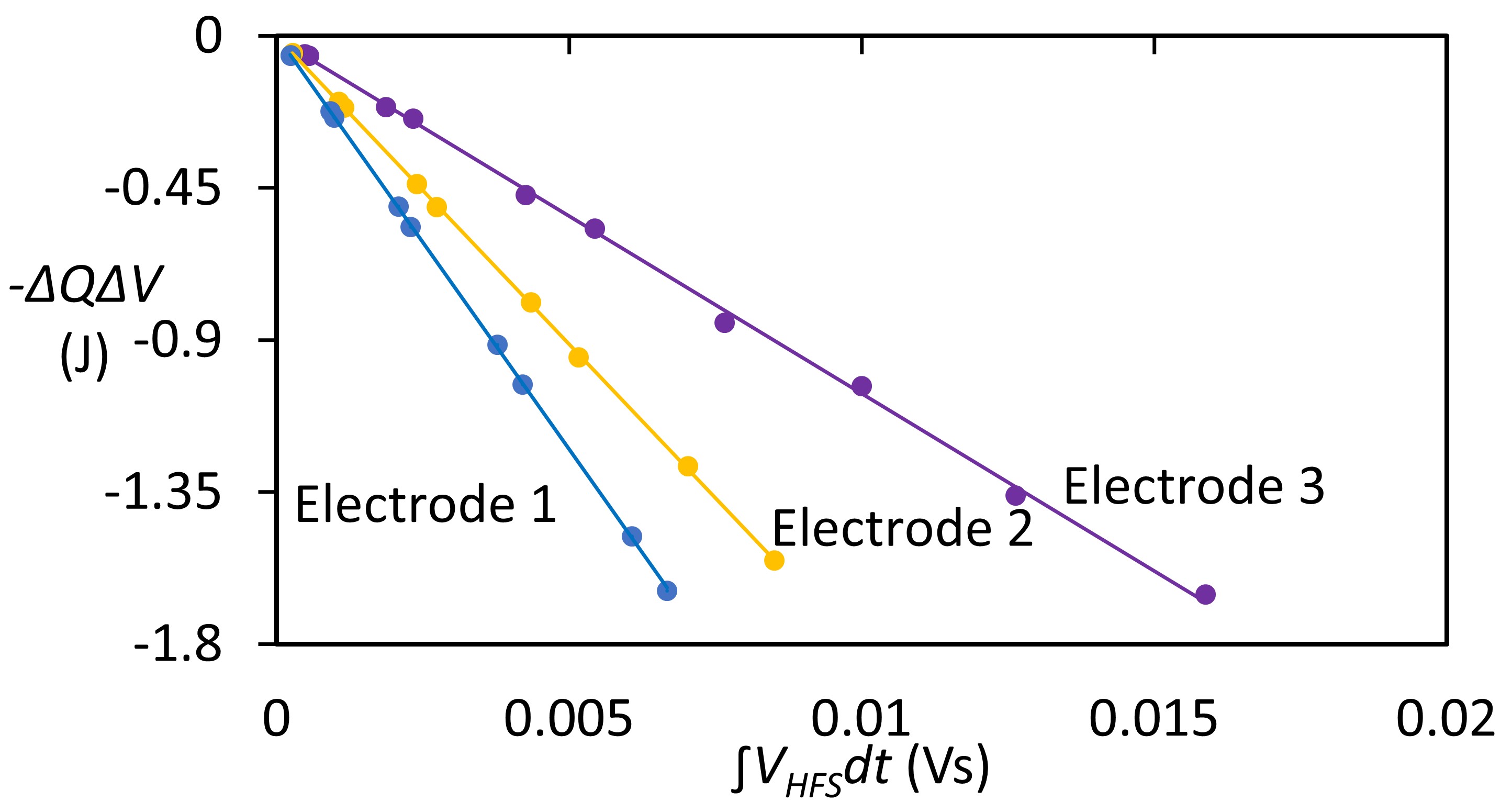}
\caption{\label{fig:calibration}Typical calibration results for 1\,M NaCl using different electrodes, with $K\textsubscript{1}$\,=\,$(-245.2 \pm 0.8)$\,JV\textsuperscript{-1}s\textsuperscript{-1}, $K\textsubscript{2}$\,=\,$(-182.1 \pm 0.6)$ JV\textsuperscript{-1}s\textsuperscript{-1}, and  $K\textsubscript{3}$\,=\,$(-104.9 \pm 1.3)$ JV\textsuperscript{-1}s\textsuperscript{-1}, the calibration constants for electrodes 1, 2, and 3, respectively.}
\end{figure}

Figure~\ref{fig:calibration} shows the effect of different electrodes on the calibration according to Eq.~\ref{eq:two}. The main difference between the electrodes is how they were mounted in the cell, affecting their heat contact with the HFS. As noted in Ref.~\onlinecite{heatmanuscript}, the calibration assumes that the HFS was equally sensitive to Joule heat and to reversible heat produced in the porous network, which was not necessarily the case. 

\begin{figure}
\includegraphics[scale=0.185]{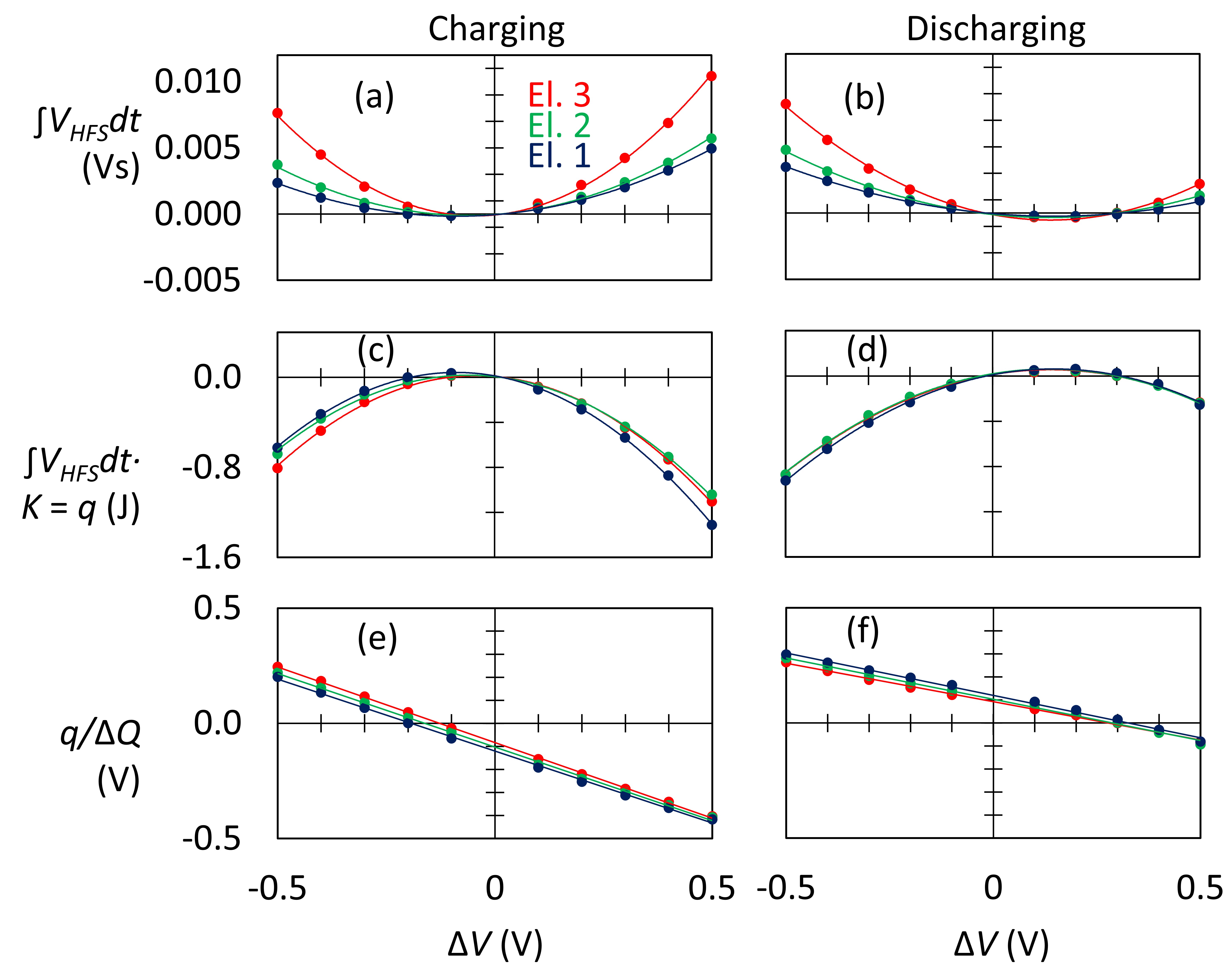}
\caption{\label{fig:results}Typical HFS measurement results during charging and discharging for different electrodes, (a-b) integrated heat flux, (c-d) calibrated heat, (e) calibrated charging heats divided by $\Delta Q$. A least squares fit to Eq.~\ref{eq:three} gives $\Delta V\textsubscript{att},=\,(-0.102 \pm 0.007)$\,V and $\left[-\frac{1}{2}-\frac{3}{2}f\right]\,=\,-0.640 \pm 0.002$. (f) Calibrated discharging heats divided by $\Delta Q$. A least squares fit to Eq.~\ref{eq:four} gives $\Delta V\textsubscript{att}\,=\,(-0.105 \pm 0.007)$\,V and $\left[\frac{1}{2}-\frac{3}{2}f\right]\,=\,-0.354 \pm 0.002$.}
\end{figure}

Figure~\ref{fig:results} shows the data analysis of HFS voltage peaks of the type in Fig.~\ref{fig:HFS}(c), with heats of charging on the left and heats of discharging on the right. Time-integrated HFS voltages are in Fig.~\ref{fig:results}(a-b). When these units are scaled using the calibration factors determined in Fig.~\ref{fig:calibration}, the results are much closer to each other (Fig.~\ref{fig:results}(c-d)).

In Ref.~\onlinecite{heatmanuscript}, formulas for the heat of charging, $q\textsubscript{ch}$, and the heat of discharging, $q\textsubscript{dis}$, were derived in terms of the electrode capacitance, the applied potential $\Delta V$, an average electric potential $f\Delta{V}$ of ions inside the pores, and a potential-independent $\Delta V\textsubscript{att}$, which corresponds to an energy per unit charge due to attraction of the ions to the electrode surface. From those equations, dividing $q\textsubscript{ch}$ and $q\textsubscript{dis}$ by the measured charge $C\Delta V$ yields the following expressions:
\begin{equation}
\frac{q\textsubscript{ch}}{C\Delta{V}}=\Delta{V}\left[-\frac{1}{2}-\frac{3}{2}f\right]+\Delta V\textsubscript{att}
\label{eq:three}
\end{equation}
\begin{equation}
\frac{q\textsubscript{dis}}{C\Delta{V}}=\Delta{V}\left[-\frac{1}{2}+\frac{3}{2}f\right]-\Delta V\textsubscript{att}.
\label{eq:four}
\end{equation}
On the basis of these two formulas, parameters $f$ and $\Delta V\textsubscript{att}$ can be obtained from linear fits of $q/(C\Delta V)$ vs. $\Delta V$, much like the linear fit of $\Delta U/(C\Delta V)$ discussed in Ref.~\onlinecite{heatmanuscript}. In Figs.~\ref{fig:results}(c) and~\ref{fig:results}(f), plots of the heat of charging and discharging, divided by $\Delta Q=C\Delta V$ indicate the same values of $\Delta V\textsubscript{att}$. When the same measurements were performed using resistance temperature detectors instead of a heat flux sensor, similar results were obtained (but with lower signal-to-noise ratio), supporting the validity of the presented calibration approach, see SI.\cite{supplemental}

\subsection{HFS sensitivity for reversible heat}
Our calibration approach assumes that the HFS has the same sensitivity for reversible heat and Joule heat, even though in practice, these two contributions to the total heat are not produced at the same location with respect to the HFS. In our test measurements, the reversible heat was fully produced in the porous network. The Joule heat, however, was generated in different resistive parts of the system. Part of the heat was generated by the silver epoxy glue contact at the side of the electrode, see Fig.~\ref{fig:cell}(c), whose resistance was $<$0.5\,\ohm. Another part of the Joule heat was generated by electrical current through the bulk solution between the Luggin tube and WE, farther away from the HFS than the porous network. This can be concluded from the concentration dependence of the total cell resistance, equal to 3.3\,$\ohm$ at 5\,M NaCl, 5.7\,$\ohm$ at 1\,M NaCl, and 20.8\,$\ohm$ at 0.1\,M NaCl, for measurements on a typical electrode. These resistances correspond to a concentration-independent contribution of about 2.5\,$\ohm$ plus the resistance of the liquid solution between Luggin tube and WE, given by $R=d/(\kappa A)$, where $1/\kappa$ is the concentration-dependent resistivity of aqueous NaCl solution obtained from Ref.~\onlinecite{Wadsworth2012}, $d\approx$\,0.6\,cm is the distance between Luggin tube and WE, and $A\approx \pi$(0.925\,cm)$^2$ is the external area of the WE exposed to bulk electrolyte solution. The thus calculated $R$ is equal to 0.7\,$\ohm$ at 5\,M, 2.2\,$\ohm$ at 1\,M, and 17.5\,$\ohm$ at 0.1\,M. Therefore, the resistance in the bulk electrolyte solution is dominant at 0.1\,M NaCl, and it is about 40\,\% of the total resistance at 1\,M NaCl and about 20\% at 5\,M NaCl. 

The calibration constant was also concentration-dependent, indicating a sensitivity of $(-5.23\pm 0.04)\times10^{-3}$\,Vs/J at 5\,M, $(-5.88\pm 0.06)\times10^{-3}$\,Vs/J at 1\,M, and $(-6.72\pm 0.02)\times10^{-3}$\,Vs/J at 0.1\,M NaCl. This can be understood in terms of where most of the Joule heat was produced. The lower the concentration, the more Joule heat was produced in bulk electrolyte solution, and the higher the sensitivity of the HFS to the Joule heat. The bulk solution is better centered with respect to the HFS than the resistive elements that give concentration-independent contributions to the total resistance, which are more to the side of the electrode, possibly not facing the HFS. 

Reversible heat produced in the porous network will flow partly toward bulk solution and partly toward the HFS. However, toward the HFS, the thermal conductivity of the glassy carbon current collector is higher (0.7-4\,W/m/K)\cite{VIEIRA2022282} than toward bulk solution, mostly consisting of water (0.6\,W/m/K)\cite{Huber2012}. This favors the flow of reversible heat in direction of the HFS. For 1\,M NaCl, in the extreme case that all the reversible heat and all the Joule heat produced in the concentration-independent 2.5\,$\ohm$ resistive elements would flow in the direction of the HFS, and only half of the Joule heat produced in the bulk electrolyte solution would flow in the direction of the HFS, the HFS would detect 75\,\% of the Joule heat and 100\,\% of the reversible heat. This would lead to an overestimation of the reversible heat by 33\,\% , since our calibration is only on the Joule heat. In reality, the flow of reversible heat and Joule heat is likely to be more evenly distributed between the directions toward and away from the HFS, resulting in a lower overestimation of the reversible heat via our calibration method.

\section{\label{sec:conclusion}conclusion}

With the presented setup, the heat of charging and discharging of a capacitive electrode can be determined as a function of electrode potential with respect to a reference electrode. In this way, differences between heat production in the anodic and cathodic ranges can be investigated.
The heat flux sensor is calibrated using the Joule heat produced in a full cycle of charging and discharging. This partly solves the problem of not knowing which fraction of electrode heat flows in the direction of the sensor.

\section{Supplementary Material}
See Supplementary material (URL will be inserted) for a 3D pdf of the electrochemical cell and a description of comparable measurements performed using an alternative setup in which resistance  temperature detectors were used instead of a heat flux sensor.

\begin{acknowledgments}
Maarten Biesheuvel is thanked for the carbon electrode material. This publication is part of the project ‘Experimental Thermodynamics of Ion Confinement in Porous Electrodes’ with project number ECHO.017.059 of the research program ECHO financed by the Dutch Research Council (NWO). 
\end{acknowledgments}

\section*{Author Declarations}
\subsection*{Conflict of Interest}
The authors have no conflicts to disclose.
\subsection*{Author Contributions}
\noindent\textbf{Joren Vos:} Formal analysis (lead); investigation (lead); methodology (supporting); supervision (equal); writing – original draft (equal); writing – review and editing (supporting). \textbf{Henrik Rodenburg:} Formal analysis (supporting); investigation (supporting); software (equal); writing – review and editing (supporting). \textbf{Danny Inder Maur:} Formal analysis (supporting); investigation (supporting); software (equal); writing – review and editing (supporting). \textbf{Ties Bakker:} Formal analysis (supporting); software (equal); writing – review and editing (supporting). \textbf{Henkjan Siekman:} Methodology (lead); writing – review and editing (supporting). \textbf{Ben Erné:} Conceptualization (lead); funding acquisition (lead); methodology (supporting); supervision (equal); writing – original draft (equal); writing – review and editing (lead). 

\section*{Data availability}
The data that support the findings of this study are available
from the corresponding author upon reasonable request.

\section*{References}
\nocite{*}
\bibliography{aipsamp}

\end{document}


\preprint{AIP/123-QED}

\title{\textbf{Supplemental Information for \\
Three-Electrode Cell Calorimeter for Electrical Double Layer Capacitors}}
\author{Joren E. Vos}
\author{Hendrik P. Rodenburg}
\author{Danny Inder Maur}
\author{Ties J. W. Bakker}
\affiliation{ 
Van ’t Hoff Laboratory for Physical and Colloid Chemistry, Debye Institute for Nanomaterials Science, Utrecht University, Padualaan 8, 3584 CH Utrecht, The Netherlands
}
\author{Henkjan Siekman}
\affiliation{\mbox{Instrumentation Department, Utrecht University, Sorbonnelaan 4, 3584 CA Utrecht, The Netherlands
}}
\author{Ben H. Erné}
\affiliation{ 
Van ’t Hoff Laboratory for Physical and Colloid Chemistry, Debye Institute for Nanomaterials Science, Utrecht University, Padualaan 8, 3584 CH Utrecht, The Netherlands
}

\date{\today}

\maketitle

\renewcommand{\thefigure}{S\arabic{figure}}

\textbf{Alternative setup with lock-in detection of RTDs using a Wheatstone bridge}

In addition to the setup described in the main text, we also developed a setup in which electrode heat is detected using a Pt100 resistance thermometer detector (RTD). Results obtained with this alternative setup support the reliability of the calibration approach applied to the HFS data in the main text. For greater sensitivity than in Ref.~\onlinecite{Janssen2017}, the RTD is connected in a Wheatstone bridge, a sinusoidal voltage is applied at high frequency, and the resulting bridge voltage is measured using a lock-in amplifier. With an ultra-thin Pt100 thermometer (TC Direct 515-113, 0.5\,mm in diameter) with its tip centered behind the working electrode, only temperature effects coming from the WE are measured.
\begin{figure}
\includegraphics[scale=0.1]{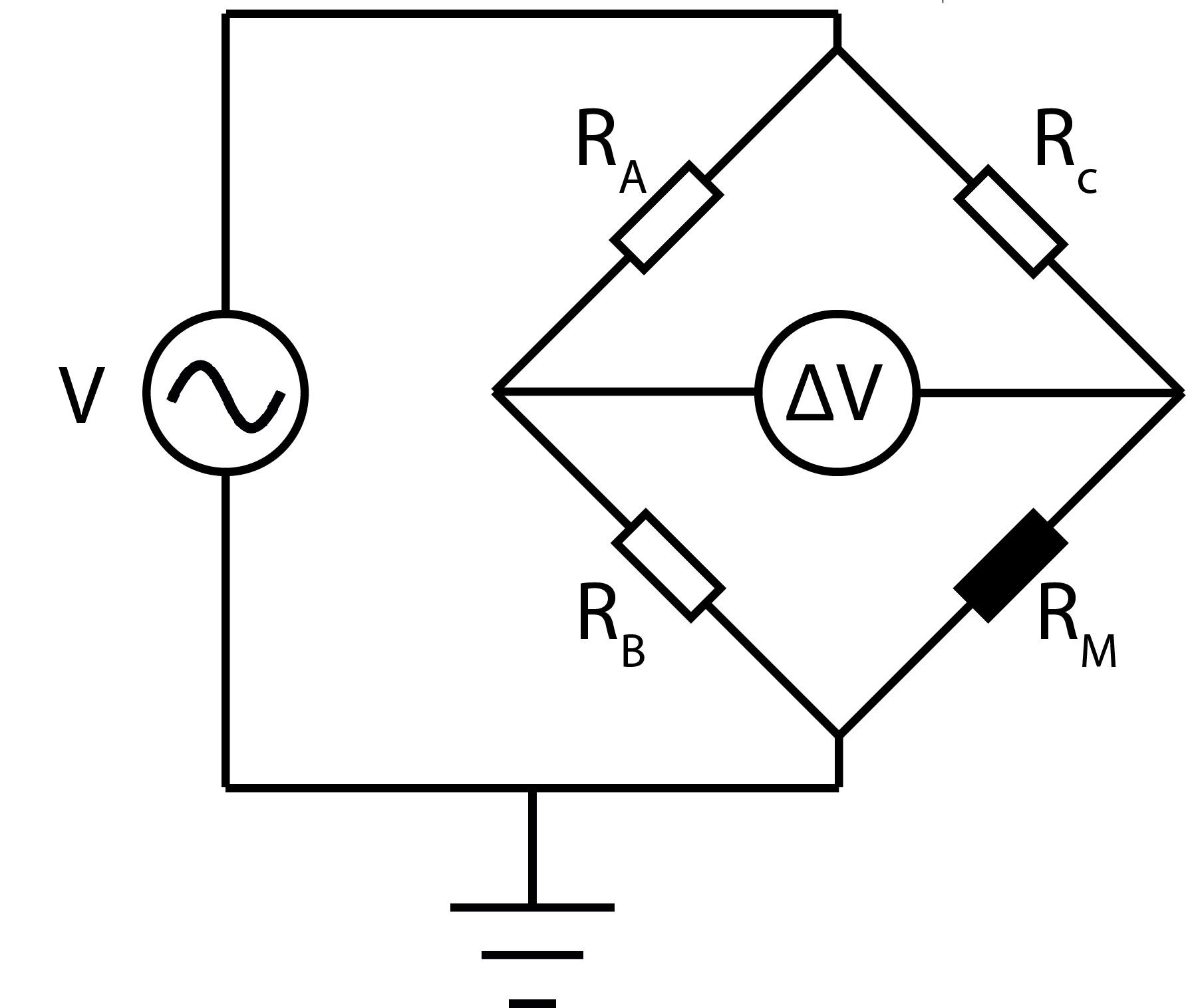}
\caption{\label{fig:Wheatstone}Four resistors connected in a Wheatstone bridge configuration, with $R\textsubscript{M}$ the Pt100 thermometer sensing the temperature of the WE and the three other Pt100 thermometers used as reference ($R\textsubscript{A}$, $R\textsubscript{B}$, $R\textsubscript{C}$). A sinusoidal voltage of 500 mV in amplitude is applied at 100 kHz, and the resulting voltage $\Delta V$ is measured with a lock-in amplifier.}
\end{figure}

The ultra-thin Pt100 thermometer ($R\textsubscript{M}$) is connected in a Wheatstone bridge configuration,\cite{Rubin1972} shown in Fig.~\ref{fig:Wheatstone}, together with three bulkier Pt100 (TC Direct 514-100) reference thermometers ($R\textsubscript{A}$, $R\textsubscript{B}$, $R\textsubscript{C}$). These three references respond more slowly to temperature fluctuations than the ultra-thin Pt100, thus contributing less noise to the measured signal.

In principle, at homogeneous temperature, the bridge is balanced. When the temperature of thermometer $R\textsubscript{M}$ changes, its resistance changes, and thus the voltage over the bridge changes. Using a Zürich Instruments HF2LI Lock-In Amplifier, a voltage of 500 mV is applied to the Wheatstone bridge at 100 kHz. The resulting voltage between the connection of $R\textsubscript{A}$ with $R\textsubscript{B}$ and the connection between $R\textsubscript{C}$ and $R\textsubscript{M}$ is measured.

The temperature difference is calculated as follows. When the bridge is not balanced, the ratio of measured voltage $\Delta V$ to applied voltage $V$ is given by $R\textsubscript{A}\approx R\textsubscript{B}\approx R\textsubscript{C}$:

\begin{equation} \tag{S1}
\label{eq:specificDeltaV}
\frac{\Delta{V}}{V}=\bigg[\frac{R\textsubscript{M}}{R\textsubscript{C}+R\textsubscript{M}}-\frac{R\textsubscript{B}}{R\textsubscript{A}+R\textsubscript{B}}\bigg]\approx\bigg[ \frac{R\textsubscript{M}}{R\textsubscript{A}+R\textsubscript{M}}-\frac{1}{2}\bigg]
\end{equation}

The voltage difference informs on the difference in resistance between the Pt100 that senses the WE temperature and that of each reference Pt100.

\begin{equation} \tag{S2}
\label{eq:s10s}
R\textsubscript{M}-R\textsubscript{A}=\frac{4\Delta{V}R\textsubscript{A}}{V-2\Delta{V}}=\Delta R
\end{equation}

The electrical resistance of the thermometers depends on the temperature: $R = 100\,\Omega + KT$, where $K$ =0.385005$\,\Omega$\text{/K} (IEC 60751 standard), thus the difference in resistance between $R\textsubscript{M}$ and $R\textsubscript{A}$ is:
\begin{equation} \tag{S3}
\label{eq:resistanceToTemperature}
\Delta R = K(T\textsubscript{M} - T\textsubscript{A}) = K\Delta T.
\end{equation}
The temperature difference is calculated by rearranging and combining Eq.~\ref{eq:resistanceToTemperature} with Eq.~\ref{eq:s10s}:
\begin{equation} \tag{S4}
\label{eq:WBridgeResult}
\Delta T =\frac{1}{K}\Delta R = \frac{1}{K}\frac{4\Delta{V}R\textsubscript{A}}{V - 2\Delta V}.
\end{equation}

This formula is an approximation of the temperature difference, in which the resistance of the leads of the RTDs is neglected, even though they are nonzero and different for the two types of RTD used ($R\textsubscript{A}$ and $R\textsubscript{M}$). This leads to an elevated baseline of $\Delta{T}$ versus time which can, however, be subtracted during analysis of the data.
\begin{figure*}
\includegraphics[scale=0.5]{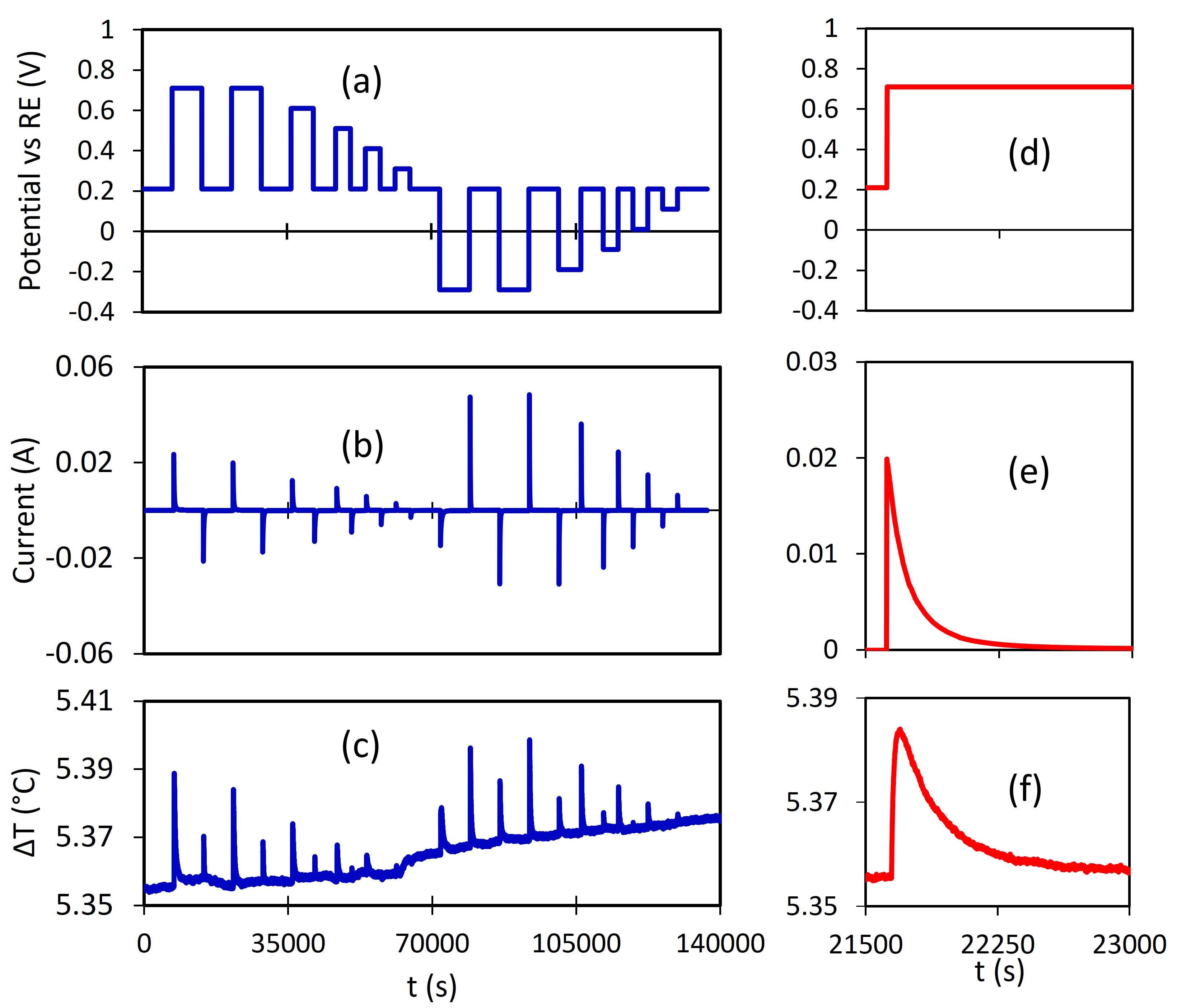}
\caption{\label{fig:results}Typical electrical and temperature difference measurements on a porous carbon electrode in 1\,M aqueous NaCl with (a) the applied potential vs RE; (b) the resulting current, and (c) the temperature difference calculated via Eq.~\ref{eq:WBridgeResult}. (d)-(f) Shape of the signals in (a)-(c), respectively, during charging from 0.21\,V to 0.71\,V (potentials versus RE; 0.21\,V is the open circuit potential).}
\end{figure*}

The calibration of the setup is comparable to the calibration in the main text, except that HFS voltage is replaced by $\Delta{T}$. In Fig.~\ref{fig:results}, which is comparable to Fig.~2 of the main text, results of a typical measurement are shown. The signal-to-noise ratio of the highest peak is 130:1 in Fig.~\ref{fig:results}, thus, more than one order of magnitude smaller than the HFS setup in the main text. The setup is more vulnerable to outside temperature fluctuations than the HFS setup and the current through the RTDs causes their heating, effects revealed by the baseline.

\begin{figure*}
\includegraphics[scale=0.35]{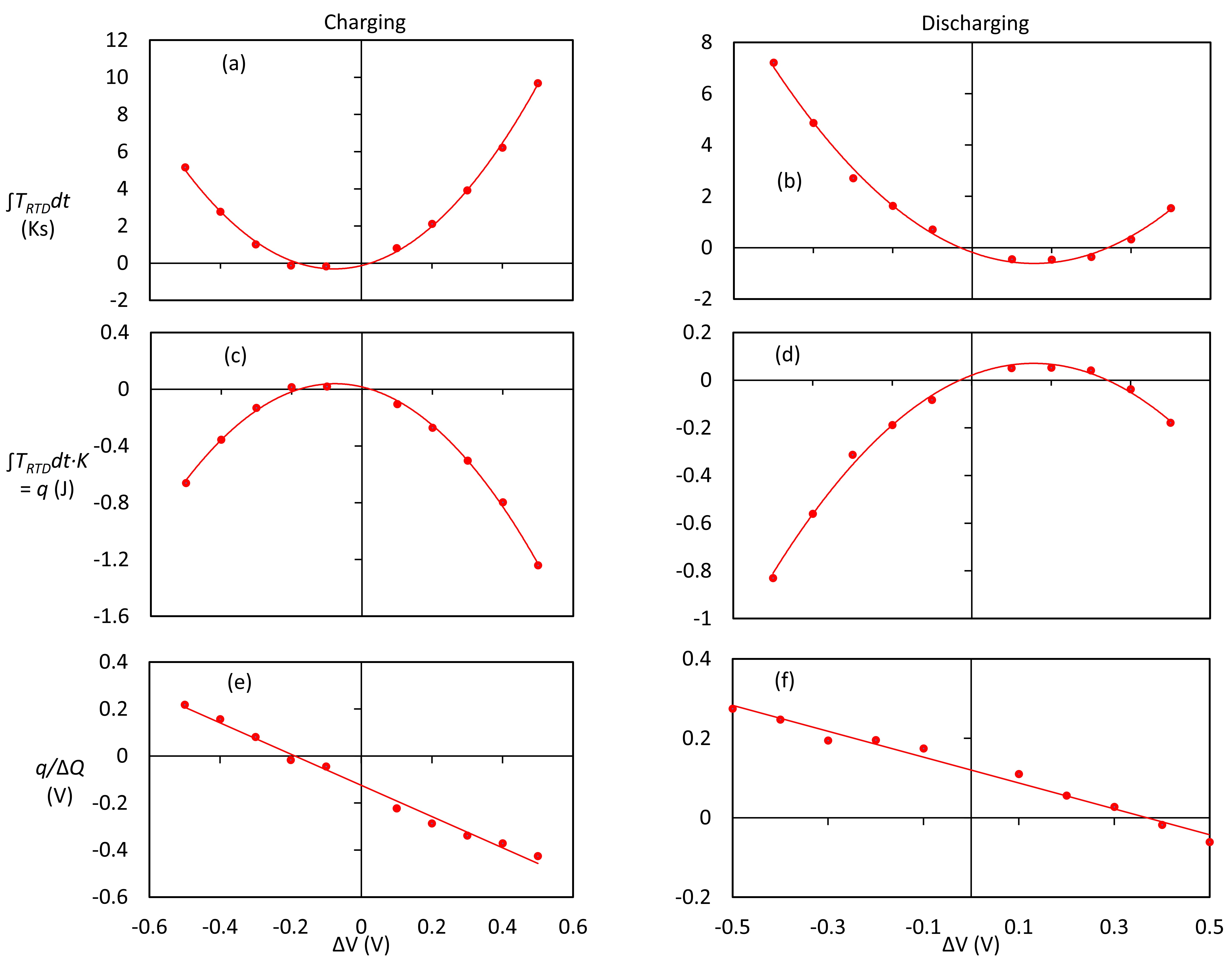}
\caption{\label{fig:calibration}Typical RTD measurement results during charging and discharging, (a-b) integrated temperature differences, (c-d) calibrated heats, (e) calibrated charging heats divided by $C\Delta{V}$. A least squares fit to Eq.~3 gives $\Delta V\textsubscript{att}=(-0.109 \pm 0.024)$\,V and $\left[-\frac{1}{2}-\frac{3}{2}f\right]=-0.660 \pm 0.008$. (f) Calibrated discharging heats divided by $C\Delta{V}$. A least squares fit to Eq.~4 gives $\Delta V\textsubscript{att}=(-0.120 \pm 0.016)$\,V and $\left[\frac{1}{2}-\frac{3}{2}f\right]=-0.326 \pm 0.005$.}
\end{figure*}

Fig.~\ref{fig:calibration} shows the results of the RTD setup in the same way as Fig.~4 of the main text. The results are comparable to those with the HFS setup presented in the main text. Nevertheless, the higher signal-to-noise ratio and the better thermostatization make the HFS a better option for the measurements.

\section*{References}
\nocite{*}
\bibliography{SIbib}